# Developing Micro DC-Brushless Motor Driver and Position Control for Fiber Positioners

Laurent Jenni[a], Philipp Hörler[a], Laleh Makarem[a], Jean-Paul Kneib[a], Denis Gillet[a], Hannes Bleuler[a], Mohamed Bouri[a], Francisco Prada[b], Guillermo De Rivera[c], Justo Sanchez[d]

[a]Ecole Polytechnique de Lausanne, LSRO-LASTRO, CH-1015, Lausanne, Switzerland

[b]Instituto de Fisica Teorica, Universidad Autonoma de Madrid, Cantoblanco, E-28049 Madrid, Spain

[c]Laboratorio de Estructura y Tecnología de Computadores II, Universidad Autonoma de Madrid, Cantoblanco, E-28049 Madrid, Spain

[d]Instituto de Astrosica de Andalucia (CSIC), Granada, E-18008, Spain

**ABSTRACT**

In the large-scale, Dark Energy Spectroscopic Instrument (DESI), thousands of fiber positioners will be used. Those are robotic positioners, with two axis, and having the size of a pen. They are tightly packed on the focal plane of the telescope. Dedicated micro-robots have been developed and they use 4mm brushless DC motors. To simplify the implementation and reduce the space occupancy, each actuator will integrate its own electronic control board. This board will be used to communicate with the central trajectory generator, manage low level control tasks and motor current feeding. In this context, we present a solution for a highly compact electronic. This electronic is composed of two layers. The first is the power stage that can drive simultaneously two brushless motors. The second one consists of a fast microcontroller and deals with different control tasks: communication, acquisition of the hall sensor signals, commutation of the motors phases, and performing position and current regulation. A set of diagnostic functions are also implemented to detect failure in the motors or the sensors, and to sense abnormal load change that may be the result of two robots colliding.

## 1. INTRODUCTION

Since 1930 we know the universe is expanding, and since a few decades astronomers have shown that expansion speed increases [1]. To explain this acceleration a new theory postulates the existence of the dark energy whose effect is antagonistic to that of the gravity force. In order to quantify its effect, high-precision measurements of universe expansion are required. Among the different observation methods one consists in using large-scale spectroscopic survey to measure the evolution of distant galaxies red-shift. This operation is carried out by placing a large number of optical fibers in the focal plane of a telescope to accurately measure the light of a large number of galactic objects.

Several projects are carried out in this direction, among which the Dark Energy Spectroscopic Instrument (*DESI*), an international project led by the *Lawrence Berkeley National Laboratory*, that aims to develop a 5000 fibers spectrograph

---

Corresponding author: philipp.horler@epfl.ch

on the *Mayall* four-meter telescope. The objective is to be able to reconfigure every 10 minutes the position of all the fiber tips in the reflected image so as to observe a new set of light dots. To do so the focal plane of the telescope is composed of 5000 robotic positioners assembled side by side in a circular 80 centimeters area (Figure 1).

**1.1 Description of the robotic positioners and their operation mode**

Each robot has two rotary axes allowing them to place the tip of one fiber within a circular surface of 1.2 centimeter (Figure 2). They are driven by miniature DC brushless motors with high ratio (1024:1) gear reducers, and they are able to change to any new position in the working area in less than 3 seconds. In order to reach all possible points in the image, the workspaces of the manipulators overlap.

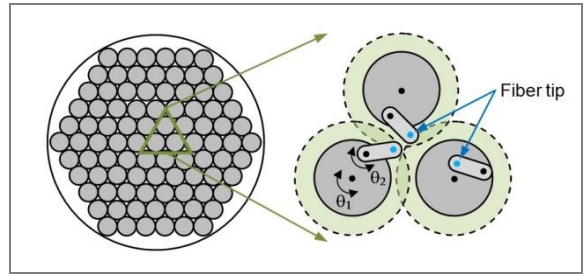

Figure 1: Positioners disposition on the focal plane of the telescope

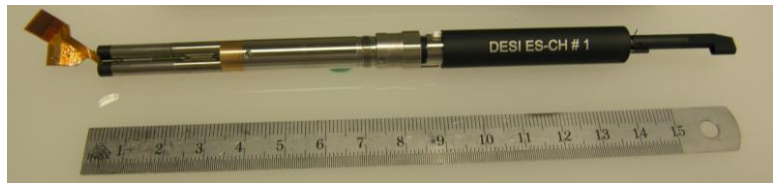

Figure 2: Prototype of the robot

The operation mode of this system is divided in two phases. First the observation phase that last for about 10 minutes and during which the robots are stopped at specific position. During this time a central unit computes the new position (and the trajectory to be followed for reaching this position) for every actuator. The second phase is the reconfiguration of the whole array for the next observation. In order to maximize the observation time the reconfiguration should be done in less than 45 seconds. Because of the large number of actuators, the position control is done in a semi decentralized way. As said above, the motions of the system are computed offline by a calculator. Since the robots can collide it allows running a complex collision avoidance algorithm [2]. But the communication between this central unit and the 5000 devices is too slow for controlling all of them simultaneously. Thus, each robot embarks its own electronic to drive and control its two motors. During the observation phase, the new trajectory is sent to every robot and when it's done the main unit broadcasts the start signal and each manipulator follows its pre stored path autonomously.

This approach leads to some difficulties. The first one is the lack of space; each actuator has to embark its own electronic for controlling and powering two axes, and this electronic should fit on a 6 x 40 millimeter board. Another issue is the regulation; the motors need to be controlled in position, but because of strict cost and space constraints, it's suitable to avoid adding coder on them. Thus the control loop has to be done only with the 60° resolution hall sensor, making it difficult to achieve stability and sufficient precision. In order to reach the requirement we propose an approach based on mixed open and closed loop control.

## 2. ELECTRONIC DESCRIPTION

The main constraint for the electronic is its size. It should fit in a 6 x 40 millimeters circuit with a maximum high of 2 millimeters. It should include a micro controller, capable to control simultaneously two axes and manage some low level task as trajectory interpolation or communication with the central unit. And it should include the power electronic for the supply of the two motors.

The main trick to save space is by using a single voltage source to drive the motor and the microcontroller. This is possible because the two brushless motors used in the robot have a low nominal voltage of 3 volts, thus a single 3.5 volts line is enough. The first advantage is that it avoids having an additional supply wire or an on board converter. The second advantage is that the micro controller can directly drive the power bridge used to supply the motors. Both transistor of a bridge used for one motor coil (Figure 3) can be fully open or closed by the microcontroller without the help of an additional driver IC. Thanks to this the power stage consists only in 6 (dual) transistors packages of 1.6 x 1.6 millimeters plus a few sub-millimeter passive components (Figure 4).

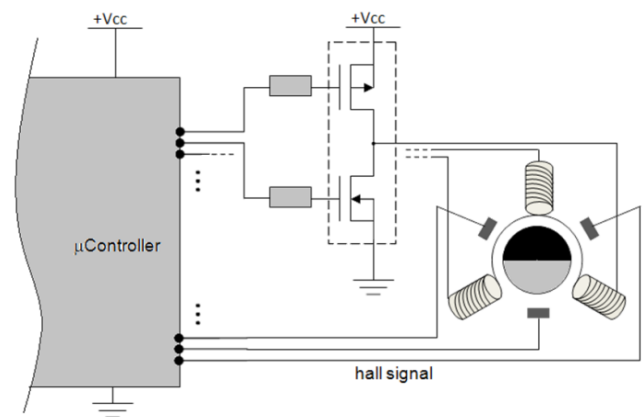

Figure 3. Direct driving of the P-N mosfet bridge by the microcontroller

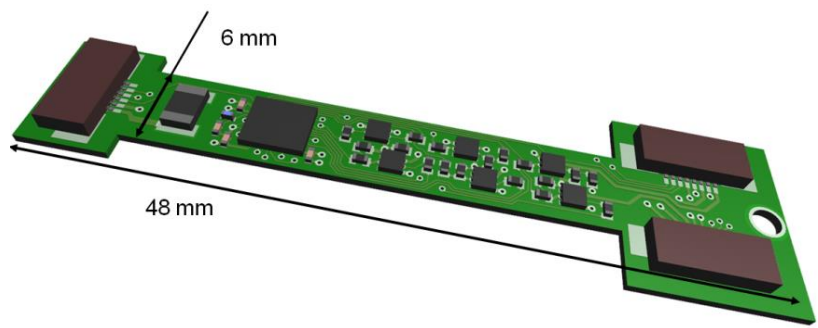

Figure 4. 3D image of the beta-version of the electronic board with non-final connectors

# 3. CONTROL ALGORITHM

The control algorithm consists in 3 principal tasks (Figure 5). The trajectory generation that stores and interpolates the desired trajectory sent by the master unit. The position loop that controls the two motors. And the blocking detection that senses abnormal torque variation on the system.

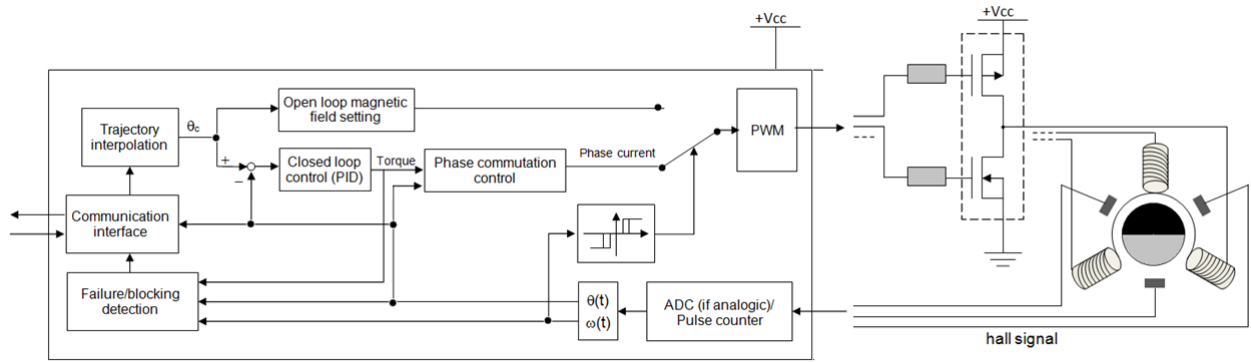

Figure 5. Diagram of the control algorithm

## 3.1 Trajectory interpolation

The communication between the central unit and the robots is done with an I$^2$C bus with 3.2MB/s data rate. In order to speed up the transmission of the 5000 trajectories they are sent under the form variable size position and time vectors. Each actuator receives two pairs of position/time vector (one per motor) and, while requested to execute the path, the controller linearly interpolates the data with a 2 kHz sampling rate in order to generate a continuous and smooth motion. With this, a robot that has no risk of collision can be feed with two single element vectors which results in a linear motion, avoiding useless data to be transmitted.

## 3.2 Position regulation

The particularity of the position control is the mixed approach of open and closed loop regulation. The open loop control consists in driving the brushless motor like a stepper motor. By forcing the proper amount of current in each of the motor phases it's possible to fix the angle of the stator magnetic field and then (if the friction is small enough) the rotor is going to align itself with this field. To do so there is no need for position measurement, but it's a very inefficient way in terms of power consumption because the motor should be feed with a lot of current in order to be sure that the rotor "sticks" to the desired consign. On the other hand, the closed loop method consists in sensing the rotor position and setting the magnetic field of the stator 90° ahead of it (or behind depending on the spinning direction). This way, the maximum amount of electrical power is converted into mechanical power, it's more efficient and the motor can run faster. Plus there is no risk of losing the position.

The closed loop method also allows to follow arbitrary trajectories. For example if a step trajectory to a new set position is applied, the PID controller will make the motor run as fast as possible to reach it. With the open loop method there is need to compute at every instant a new position of the magnetic field that takes into account the limits in terms of speed and acceleration of the system, otherwise the rotor will not follow it. Typically, the linear interpolation described in *§3.1* would not be possible in open loop.

The necessity of combining this two methods, comes from the lack of a proper position sensor (like an optical encoder). The only position measurement is retrieved by the 3 digital hall sensors with a resolution of 60°. This low resolution is a problem for the accuracy, as well as the stability of the system. For the precision, and since there is a 2000:1 reduction ratio between the motor and the system output, a 60° rotation in one motor corresponds to a displacement of up to 4 micrometers of the optical fiber. This is 2 times the resolution defined in the specification (2μm). With the open loop control, the positioning resolution is proportional to the resolution of the current supply (in this case $< 0.1°$).

The closed loop stability issue appears at low speed. Instead of a smooth motion the motor will performs a succession of 60° jumps followed by a short stop. It induces vibration that are not only problematic because is speeds up the wearing of the mechanic, but also for the precision because it makes it impossible to apply backlash catch up methods.

### 3.3 Collision detection

As said above, for being able to place a fiber tip anywhere in the focal plane of the telescope, the workspaces of the manipulators overlap. The trajectories are computed in order to avoid collision between the actuators, but if an error occurs in the position measurement (for example) and two robots collide, the torque is sufficient to damage them. Thus it's mandatory to rapidly detect abnormal load variations to avoid damage when a collision occurs. This can be done in closed loop regulation because the controller is going to feed the motor with enough current to follow the desired trajectory and as soon as the system runs at a constant speed the current depends only on the torque load which is considered to be constant. Thus a rapid variation of the current can only be interpreted as a collision. The drawback is that the current variation relies on the position measurement, and thus can be (at best) detected only after a 60° step of the motor. Then, the mechanical parts used to link the motor and the output of the system, have to be flexible enough to sustain this deformation.

## 4. RESULTS

### 4.1 Tracking error and current consumption

The results in figures 7 and 8 are obtained with a test trajectory of a trapezoidal speed profile shown in figure 6. In figure 7 we can observe that the closed loop control has a better tracking of the position at high speed, whereas the open loop control has an error proportional to the speed. At standstill however, the closed loop control is unstable. Because the set position is in between two increments of the hall sensors, the set position can never be achieved, and thus the system starts oscillating around the set position.

In figure 8 we can observe that the open loop control is consuming more current than the closed loop control. The closed loop control gives just enough current in order to move the rotor, whereas in the open loop control we have to set the current higher to ensure that we don't lose any steps. In this case the current was set constant. All the exceeding current which is not transformed into mechanical torque is generating heat dissipation in the motor coils.

### 4.2 Collision detection

Since there were not enough built prototypes to take the risk of damaging one of them, the following results are done by simulation and are based on an approximated value for stiffness of the system. The simulation is done with *Simulink* (from *MATLAB*), the model includes the motor (electronic and mechanic model) and a simplified version of the cinematic chain in the actuator. It further includes the control algorithm as it is implemented in the microcontroller of the board, and it includes a model of the power bridge electronics. The case shown in the figures 9 and 10 corresponds to the robot with one motor running at a constant speed of 500rpm and hitting a hard stop. The rapid current variation is sensed by the regulator and after 6 milliseconds the driver is turned off.

## 5. CONCLUSION

In this paper, a very compact and power efficient electronic board for controlling a two degrees of freedom robotic positioner is presented. Different innovative approaches result in a small number of electronic components and a very power efficient and precise control of two brushless DC motors. These approaches can be summarized as follows:

- The use of the same voltage for the microcontroller and the motors
- The simplified power bridge using two mosfet which can be commuted directly by the microcontroller
- The combination of closed loop and open loop control strategies, taking the advantages of both methods
- The real time trajectory interpolation

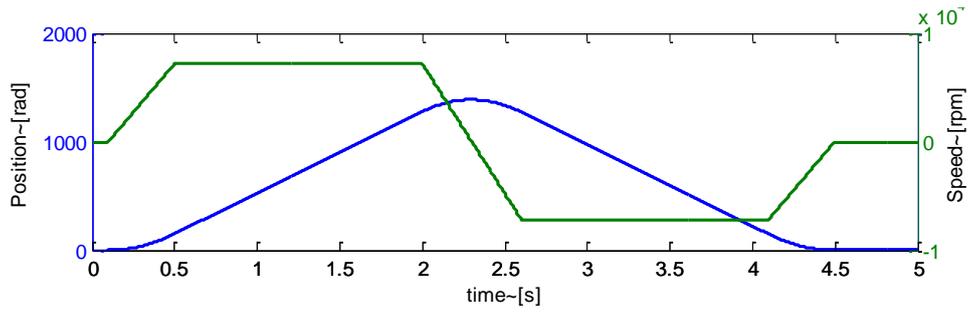

Figure 6. Test trajectory

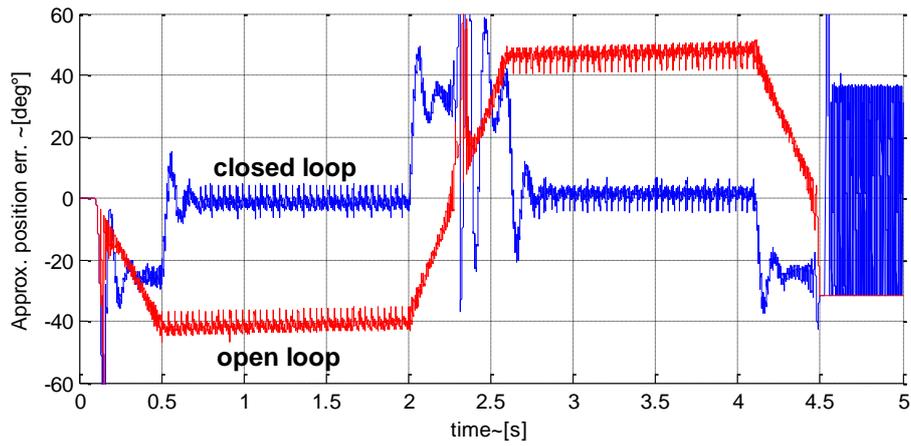

Figure 7. Tracking error

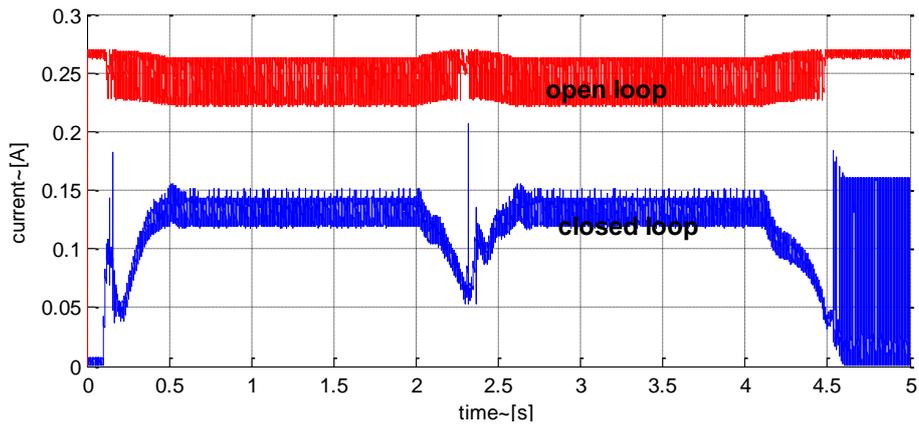

Figure 8. Current consumption

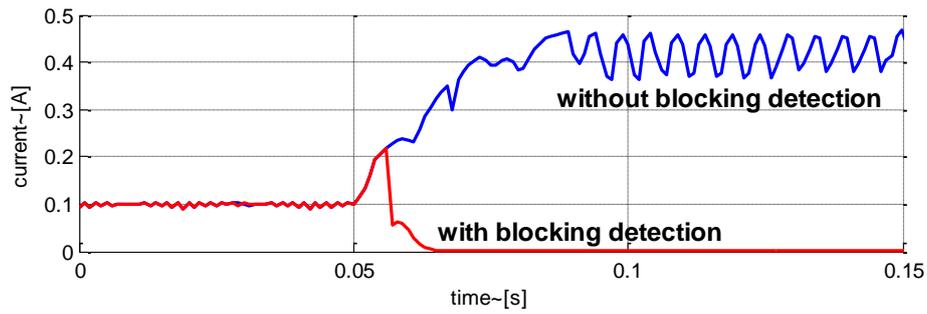

Figure 9. Evolution of the current after the hard stop

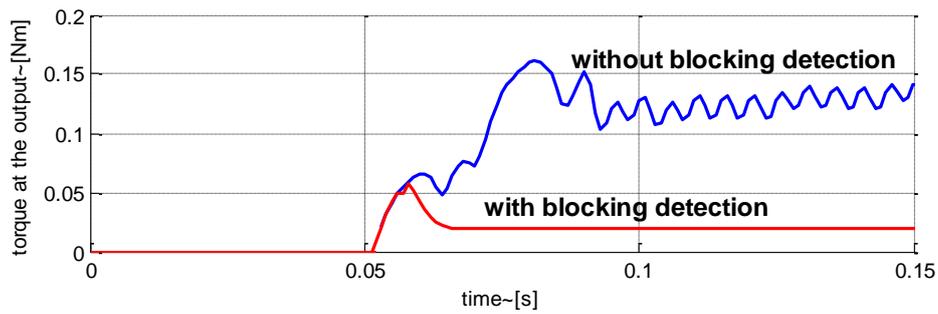

Figure 10. Evolution of the torque after the hard stop at the system output